\documentclass[aps,onecolumn,nofootinbib,superscriptaddress,floatfix]{revtex4}

\usepackage{color,amsmath,amssymb,graphicx,latexsym,subfigure}

\usepackage{multirow}
\usepackage{ulem}
\usepackage{subcaption}

\usepackage{float}
\usepackage{booktabs}
\usepackage{threeparttable}
\usepackage[colorlinks,linkcolor=green, anchorcolor=green,citecolor=green]{hyperref}
\usepackage{ulem,soul}
\usepackage{xcolor}

\newcommand{\be}{\begin{equation}}
\newcommand{\ee}{\end{equation}}
\newcommand{\ba}{\begin{eqnarray}}
\newcommand{\ea}{\end{eqnarray}}

\newcommand{\gsim}{\mathrel{\hbox{\rlap{\lower.55ex \hbox {$\sim$}}
			\kern-.3em \raise.4ex \hbox{$>$}}}}
\newcommand{\lsim}{\mathrel{\hbox{\rlap{\lower.55ex \hbox {$\sim$}}
			\kern-.3em \raise.4ex \hbox{$<$}}}}

\makeatletter
\def\section{\@startsection {section}{1}{\z@}%
    {-3.5ex \@plus -1ex \@minus -.2ex}%
    {2.3ex \@plus.2ex}%
    {\normalfont\bfseries\boldmath\rightskip\z@}}
\makeatother

\hypersetup{colorlinks=true,
	breaklinks=true,
	pdfstartview=Fit,
	linkcolor=blue,
	citecolor=blue,
	urlcolor=blue}

\begin{document}

\title{The Quintom theory of dark energy after DESI DR2}

\author{Yifu Cai} \email{yifucai@mail.ustc.edu.cn}
\affiliation{Department of Astronomy, School of Physical Sciences,
 University of Science and Technology of China, Hefei 230026, China}
\affiliation{CAS Key Laboratory for Research in Galaxies and Cosmology,
 School of Astronomy and Space Science, University of Science and Technology of China, Hefei 230026, China}
 
\author{Xin Ren} \email{rx76@ustc.edu.cn}
\affiliation{Department of Astronomy, School of Physical Sciences,
 University of Science and Technology of China, Hefei 230026, China}
\affiliation{CAS Key Laboratory for Research in Galaxies and Cosmology,
 School of Astronomy and Space Science, University of Science and Technology of China, Hefei 230026, China}
 
\author{Taotao Qiu} \email{qiutt@hust.edu.cn}
\affiliation{School of Physics, Huazhong University of Science and Technology, Wuhan, 430074, China}

\author{Mingzhe Li} \email{limz@ustc.edu.cn}
\affiliation{Interdisciplinary Center for Theoretical Study \& Peng Huanwu Center for Fundamental Theory, University of Science and Technology of China,  Hefei, Anhui 230026, China}
 
\author{Xinmin Zhang} \email{xmzhang@ihep.ac.cn}
\affiliation{Theoretical Physics Division, Institute of High Energy Physics, Chinese Academy of Sciences, 19B Yuquan Road, Shijingshan District, Beijing 100049, China}
\affiliation{School of Nuclear Science and Technology, University of Chinese Academy of Sciences, Beijing,101408, China}

\begin{abstract}
Observations from The Dark Energy Spectroscopic Instrument Data Release 2 (DESI DR2) are challenging the $\Lambda$ Cold Dark Matter ($\Lambda$CDM) paradigm by suggesting that the equation-of-state parameter of dark energy evolves across $w = -1$, a phenomenon known as the Quintom scenario. Inspired by this development, we present a staged review of Quintom cosmology including its theoretical foundations, observational supports, and implications as well as possible extensions. We first trace the historical progression from Einstein’s static cosmological constant to modern dynamical dark energy, summarizing recent cosmological constraints that favor an evolving $w(z)$ along time. A key focus is the theoretical no-go theorem for dark energy showing that no single canonical field or perfect fluid model can smoothly cross the $w = -1$ boundary. We then survey viable Quintom constructions, including two-field models, single-scalar fields with higher derivatives, modified gravity frameworks, interacting dark energy, and an effective field theory approach that unifies these mechanisms. Possible interactions of Quintom fields with ordinary matter and the potential roles in yielding non-singular universe solutions are discussed.
\end{abstract}
 
\maketitle

\section{INTRODUCTION}

The accelerated expansion of the Universe was discovered in 1998 through measurements of distances from high-redshift Type Ia supernovae (SNe) \cite{SupernovaSearchTeam:1998fmf, SupernovaCosmologyProject:1998vns}, further confirmed by the Cosmic Microwave Background (CMB) and other cosmological observations. The simplest explanation, a constant dark energy $\Lambda$, was then introduced to describe this acceleration, forming the basis of the standard $\Lambda$CDM cosmological scenario.

The Dark Energy Spectroscopic Instrument (DESI) is a state-of-the-art astronomical instrument designed to conduct groundbreaking studies of the dark universe~\cite{DESI:2016fyo,DESI:2022xcl,DESI:2024uvr,DESI:2024lzq}. Installed at the Kitt Peak National Observatory's Mayall telescope, DESI's primary mission is to create a detailed 3D map of the universe by measuring the spectra of more than 30 million galaxies and quasars. By analyzing these spectra, scientists aim to shed light on the nature of dark energy, the mysterious force driving the accelerated expansion of the universe. 

More recently, measurements of baryon acoustic oscillations (BAO) from DESI 2024, when combined with SNe datasets, have provided evidence for dynamical dark energy at a confidence level of $2.5$--$3.9\sigma$ \cite{DESI:2024mwx}. In particular, the DESI 2024 data exhibit a preference for a Quintom behavior \cite{Feng:2004ad}, wherein the equation of state (EoS) parameter of dark energy, $w$, crosses the cosmological constant boundary ($w = -1$). This dynamic evolution of dark energy has garnered significant attention and stimulated extensive investigation.

The latest version of DESI Data Release 2, when analyzed together with SNe constraints, further reinforces this preference, increasing the statistical significance to more than $4\sigma$ \cite{DESI:2025zgx, DESI:2025fii, DESI:2025wyn}. These findings have motivated renewed interest in exploring models of dynamical dark energy that can accommodate such behavior, highlighting the need for further theoretical studies of Quintom dark energy~\cite{Li:2024qso,Cortes:2024lgw, DESI:2024aqx, DESI:2024kob, Yang:2024kdo, Wang:2024dka, Carloni:2024zpl, Giare:2024gpk, Mukherjee:2024ryz, Jiang:2024xnu, Dinda:2024kjf, Yang:2025kgc, Giare:2024oil, Liu:2024gfy, Escamilla-Rivera:2024sae, Yin:2024hba, Chudaykin:2024gol,Huang:2025som, RoyChoudhury:2024wri, Odintsov:2024woi, Ormondroyd:2025iaf, Pang:2025lvh, Anchordoqui:2025fgz, Pan:2025qwy, Pan:2025psn, Paliathanasis:2025dcr, Abedin:2025yru, Odintsov:2025kyw,Cheng:2025lod,Sadeghi:2017nxe}.

This article is organized as follows: Section~\ref{sec:2} traces the development from Einstein’s cosmological constant to modern observational evidence for an evolving dark‐energy component, motivating a phenomenological classification of equations of state. Section~\ref{sec:3} confronts these possibilities with the latest high-precision datasets—DESI BAO, supernovae and CMB—showing a growing preference for a Quintom-B scenario.  Section~\ref{sec:4} reviews the No-Go theorem that forbids $w=-1$ crossings in single‐field or single perfect‐fluid models, thereby setting the theoretical stage for viable Quintom constructions. Section~\ref{sec:5} surveys concrete model‐building avenues including multi‐scalar systems, scalar fields with higher derivatives, modified gravity realizations, interacting dark energy, and an effective-field-theory unification. Section~\ref{sec:6} considers the implications of Quintom cosmology from two perspectives. One is to explore possible couplings of Quintom dark energy to ordinary matter, outlining the resulting laboratory and cosmological signatures. The second is to discuss the role of Quintom dynamics in non-singular early-universe scenarios such as bounce, cyclic and emergent solutions. Section~\ref{conclusion} is a brief conclusion.

\section{FROM CONSTANT TO DYNAMICS}
\label{sec:2}

In 1917, Albert Einstein introduced a cosmological constant, $\Lambda$, in the development of General Relativity \cite{1917SPAW}. Before 1998, it was widely considered that the cosmic expansion would clump due to gravitational attraction, and a deceleration parameter $q$ was introduced to quantify this expected behavior. Surprisingly, cosmological measurements of distant supernovae revealed an unexpected negative value for $q$, indicating that the expansion of the universe is not slowing down, but rather accelerating. This discovery was later corroborated by independent probes, including observations of the CMB and large-scale structure (LSS), and then naturally led to a profound fact: our universe is accelerating right now. To address this, the concept of dark energy, which can be characterized as a smooth, spatially homogeneous energy component with negative pressure, dominating the universe at present on large scales, was proposed.

By introducing a constant term $\Lambda$ into General Relativity to account for the observed acceleration of the universe, the $\Lambda$CDM model provides a simple and empirically successful framework for modern cosmology. It assumes that General Relativity is the correct theory in describing gravity on cosmological scales and the Universe consists of radiation, ordinary matter, cold dark matter (CDM) and a cosmological constant $\Lambda$ which is associated with dark energy whose density remains constant even in an expanding background.

Within the framework of the $\Lambda$CDM cosmology, the evolution of the homogeneous and isotropic background universe is governed by the Friedmann equations:
\begin{equation}\begin{aligned}
 & H^2\equiv\left(\frac{\dot{a}}{a}\right)^2=\frac{8\pi G\rho}{3}, \\
 & \frac{\ddot{a}}{a}=-\frac{4\pi G}{3}\left(\rho+3p\right),
\end{aligned}\end{equation}
where $H$ is the Hubble parameter, $\rho$ and $p$ denote the total energy density and pressure of all components in the universe at a given time. The acceleration of the universe requires $\ddot{a}>0$, which implies an effective EoS parameter $w = p/\rho < - 1/3$. The cosmological constant $\Lambda$ can thus be interpreted as a dark energy component with a constant equation of state $w_{\Lambda} = -1$.

Despite the remarkable success, the introduction of the cosmological constant $\Lambda$ presents some profound theoretical challenges~\cite{Martin:2005bp,Linder:2007ka, Frieman:2008sn}. One is the famous cosmological constant problem: quantum mechanically, field theories typically expect a vacuum energy density that is more than 120 orders of magnitude larger than the observed value, $\rho_{\Lambda}^{\mathrm{obs}} \sim (10^{-3}\mathrm{eV})^{4}$. This mechanism also suffers from the cosmic coincidence problem: it is puzzling why the energy density of dark energy is of the same order of magnitude as that of matter precisely at the present epoch.

Except for the aforementioned conceptual issues, several observational tensions or anomalies have emerged, especially in recent years \cite{Nesseris:2006jc,Wang:2018fng,Heisenberg:2022gqk,Heisenberg:2022lob,DiValentino:2021izs,Perivolaropoulos:2021jda,Abdalla:2022yfr,Lee:2022cyh,CosmoVerseNetwork:2025alb}, including the Hubble tension, the $\sigma_8$ tension, large-angle anomalies in the CMB, a possible cosmic dipole, etc. These discrepancies indicate that the prevailing $\Lambda$CDM model may be incomplete and motivate the exploration of its extensions. In the literature, a lot of studies have been performed to address part of these challenges, among which dynamic models of dark energy, whose $w$ is not always equal to $ -1$, have attracted much attention\cite{Copeland:2006wr,Dymnikova:2000gnk,Mukhopadhyay:2007ed}.

In general, dark energy models can be phenomenologically categorized into the following main classes:
\begin{itemize}
  \item $w = -1$: This corresponds to the cosmological constant $\Lambda$.
  \item $w > -1$: The EoS lies above the cosmological constant boundary, usually called quintessence dark energy.
  \item $w < -1$: The EoS lies below the cosmological constant boundary, usually called phantom dark energy.
  \item $w$ crosses $-1$: The EoS can evolve across the cosmological constant boundary, usually dubbed Quintom dark energy. If the crossing occurs from above to below with time, it is known as Quintom-A; if the crossing is from below to above, it is known as Quintom-B.
\end{itemize}

Different values of $w$ also introduce distinct theoretical challenges. For instance, when $w < -1$, the null energy condition (NEC) is commonly violated \cite{Hawking:1973uf}, making phantom energy difficult to realize with ordinary matter fields, and can also be avoided in certain circumstances \cite{Qiu:2007fd,Moghtaderi:2025cns}.

Given the lack of a fundamental theory that naturally explains the origin of dark energy, phenomenologists often adopt a pragmatic approach by constructing models based on its macroscopic behavior and constraining the relevant parameters using observational data. A common example is the $w$CDM model, which assumes a constant but arbitrary EoS $w\neq-1$. A more general class is the $w_0w_a$CDM model, where the evolution of $w$ over time is described by two parameters. One commonly used parameterizations for dynamical dark energy is the Chevallier–Polarski–Linder (CPL) parameterization \cite{Chevallier:2000qy, Linder:2002et}, which expresses $w(z)$ as
\begin{equation}
 w(z)=w_0+w_a(1-a).
\end{equation}

With the advent of precision cosmology, during 2003 to 2012, WMAP provided increasingly stringent constraints on both the $w$CDM and the $w_0w_a$CDM models. Yet, these analyses did not show significant deviations from the $\Lambda$CDM model, as summarized in Table~\ref{WMAP}. When deriving cosmological parameter constraints, WMAP results were frequently combined with additional datasets, including BAO measurements from surveys such as 2dFGRS \cite{2dFGRS:2001csf} and 6dFGS \cite{Beutler2011The6G}, as well as SN data sets such as SNLS \cite{SNLS:2005qlf} and Union2 \cite{Suzuki2011THEHS}, in addition to small-scale CMB data and independent measurements of the Hubble constant $H_0$. 

\begin{table*}
  \centering
  \caption{Developments of the observation constraints on the dark energy EoS from WMAP 2003-2012.}
  \label{WMAP}
  \begin{tabular}{cccc}
    \hline
    data                             & $w$                          & $w_0$  & $w_a$  \\
    \hline
    WMAP1 \cite{WMAP:2003elm}          ~&~ $-\,0.98\pm 0.12$           ~&~  - ~&~  -\\
    WMAP3 \cite{WMAP:2006bqn}          ~&~ $-\,0.967{}^{+\,0.073}_{-\,0.072}$ ~&~   - ~&~  -\\
    WMAP5 \cite{WMAP:2008lyn}          ~&~ $-\,1^{+0.12}_{-0.14}$       ~&~   $-\,1.06\pm 0.14$ ~&~  $0.36\pm 0.62$ \\
    WMAP7 \cite{Komatsu2010SEVENYEARWM} ~&~ $-\,1.10\pm 0.14$            ~&~  $-\,0.93\pm 0.13$ ~&~  $w = -\,0.41^{+\,0.72}_{-\,0.71}$ \\
    WMAP9 \cite{Bennett2012NINEYEARWM}  ~&~ $-\,1.073^{+0.090}_{-0.089}$ ~&~   - ~&~  -\\
    \hline
  \end{tabular}
\end{table*}

In 2013, Planck released its first-year data \cite{Planck:2013pxb}, showing $w= -1.13^{+0.13}_{-0.14}$  (Planck2013 + WMAP9 + SNLS), which lies within $2\sigma$ of the phantom regime. Subsequent improvements in supernova calibration led to the Joint Light-curve Analysis (JLA) dataset released in 2014 \cite{SDSS:2014iwm}, which significantly enhanced the spectral calibration of SNLS data and exhibited a $1.8\sigma$ discrepancy compared to SNLS-3. Incorporating JLA data with Planck2013 results brought the constraints on dark energy back in line with 
$\Lambda$CDM. In Planck2015 \cite{Planck:2015fie,Planck:2015bue}, which adopted JLA as its default SN dataset, the previous SNLS-induced deviation from the standard model was similarly resolved. The final result is $w = -1.006^{+0.085}_{-0.091}$.

Moreover, the emergence of new observational data sets including DESY1 \cite{DES:2017qwj, DES:2017myr}, Pantheon \cite{Pan-STARRS1:2017jku}, Planck2018 \cite{Planck:2018vyg}, and DESY3 \cite{DES:2021wwk, DES:2022ccp}  further constrained the $w$CDM and $w_0w_a$CDM models~\cite{Park:2024pew}. In all of these studies, the joint analyses have continued to show no significant deviation of $w$ from $-1$. 
In 2022, the Pantheon+ supernova compilation \cite{Brout:2022vxf} reported $ w= -0.90\pm 0.14$ (SN only) and $w= -0.978^{+0.024}_{-0.031}$  when combined with CMB and BAO data. Although the fit to $w_0$ and $w_a$  remained consistent with $\Lambda$CDM within the $2\sigma$ confidence level, the results showed a slight deviation compared to previous data sets.
Meanwhile, this trend was further supported by the Union3 compilation~\cite{Rubin:2023jdq}, which yielded constraints on the $w_0w_a$CDM model showing the mild tension with $\Lambda$CDM at the $1.7-2.6\sigma$, which favored models with $w_0\textgreater -1$ and $w_a\textless 0$.
These results point to an evolving dark energy component whose EoS increases over time, suggesting a present value $w\textgreater -1$.

In 2024, following the hints of dynamical dark energy revealed by the Pantheon+ and Union3 compilations, DESY5 shows that, whether using supernova data alone or in combination with CMB, BAO, and $3\times 2pt$ measurements, the best-fit values of $w$ are consistently slightly greater than $-1$ at more than the $1\sigma$ level. These findings are in agreement with those obtained from the Union3 compilation, further supporting the trend toward a mildly dynamical dark energy. In the same year, DESI released its first year data of BAO measurements \cite{DESI:2024mwx}, which yield constraints on $w_0$ and $w_a$ that deviate from the standard model at the levels of $2.6\sigma$, $2.5\sigma$, $3.5\sigma$ and $3.9\sigma$,  when combined with data from the CMB, Pantheon+, Union3, and DESY5, respectively. These results favor a dynamical dark energy scenario characterized by $w_0 \textgreater -1, w_a \textless 0, w_0+w_a\textless -1$, i.e. Quintom-B scenario. 

The DESI DR2 of the BAO measurements \cite{DESI:2025zgx} combining with CMB alone {prefer $w_0>-1$ and $w_a<0$ over the $\Lambda$CDM model at a significance level of $3.1\sigma$ in the CPL parameterization suggesting a evolving dark energy}. When combined with Pantheon+, Union3, and DESY5, {the significance of preference for a dynamical dark energy model reaches $2.8\sigma,3.8\sigma$ and $4.2\sigma$ in the CPL parameterization, respectively.} Compared to DR1, DR2 shows improved precision and reduced uncertainties. Following the release of DESI BAO DR2, \cite{DESI:2025fii} conducted an extended analysis on the behavior of dark energy, confirming the evidence for dynamical dark energy.

Researchers are eager to uncover the potential new physics underlying these discrepancies and to explore their fundamental nature. In order to realize the dynamical behavior of dark energy, a variety of theoretical models have been proposed. Representative scalar-field models include quintessence \cite{Ratra:1987rm,Wetterich:1987fm}, phantom \cite{Caldwell:1999ew}, Quintom~\cite{Feng:2004ad,Feng:2004ff,Guo:2004fq,Hu:2004kh, Elizalde:2004mq}, k-essence~\cite{Chiba:1999ka, Armendariz-Picon:2000nqq}, and so on. In addition, some theories attribute the driving force of cosmic acceleration to the modification of General Relativity, attempting to reproduce the accelerated expansion from the geometric structure of gravity. 


\section{QUINTOM DARK ENERGY UNDER LATEST OBSERVATIONS}
\label{sec:3}

\begin{figure*}[htbp]
    \centering
        \includegraphics[width=\textwidth]{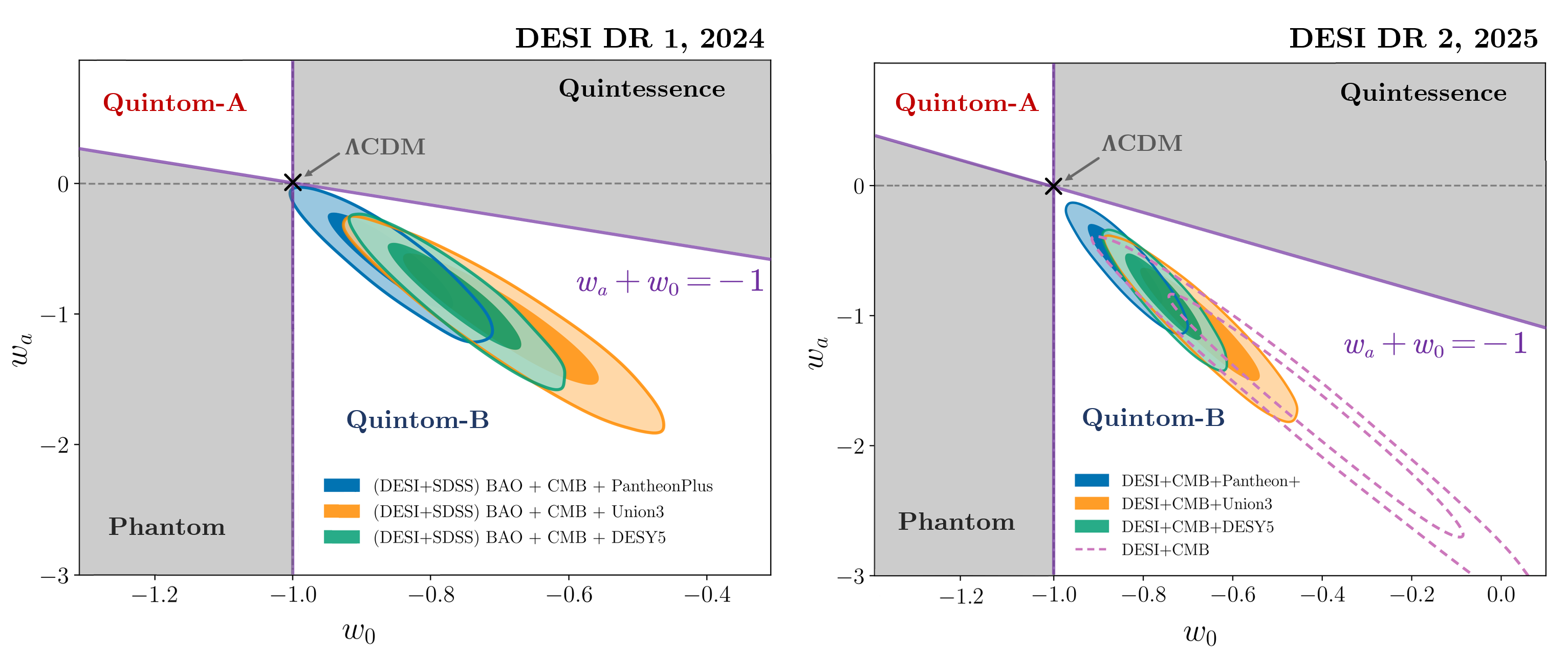}
    \caption{\it{68\% and 95\% marginalised posterior constraints on $w_0$-$w_a$ plane for the flat $w_0w_a$ model from the combination of DESI BAO data, CMB and SNe, for PantheonPlus \cite{Brout:2022vxf}, Union3 \cite{Rubin:2023jdq} and DESY5 \cite{Abbott:2024agi} SNe datasets in blue, orange and green, respectively. The left panel is taken from ref.~\cite{DESI:2024mwx}, while the right panel is from ref.~\cite{DESI:2025zgx}. We have added the boundary lines to illustrate the regions of quintessence, phantom and quintom.}}
    \label{fig:desi_w0wa_all}
\end{figure*}

\begin{figure*}[htbp]
    \centering
    \begin{minipage}{0.45\textwidth}
        \includegraphics[width=\textwidth]{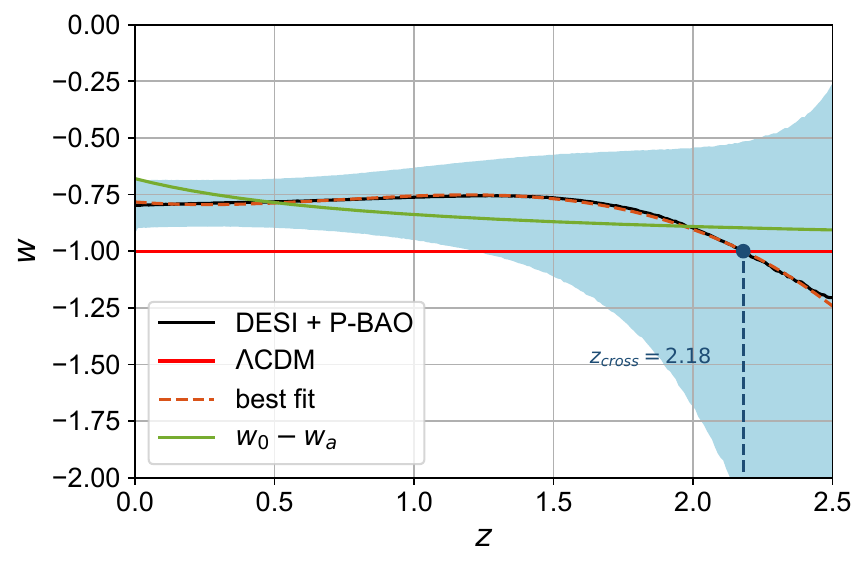}
    \end{minipage}
    \hfill
    \begin{minipage}{0.5\textwidth}
        \includegraphics[width=\textwidth]{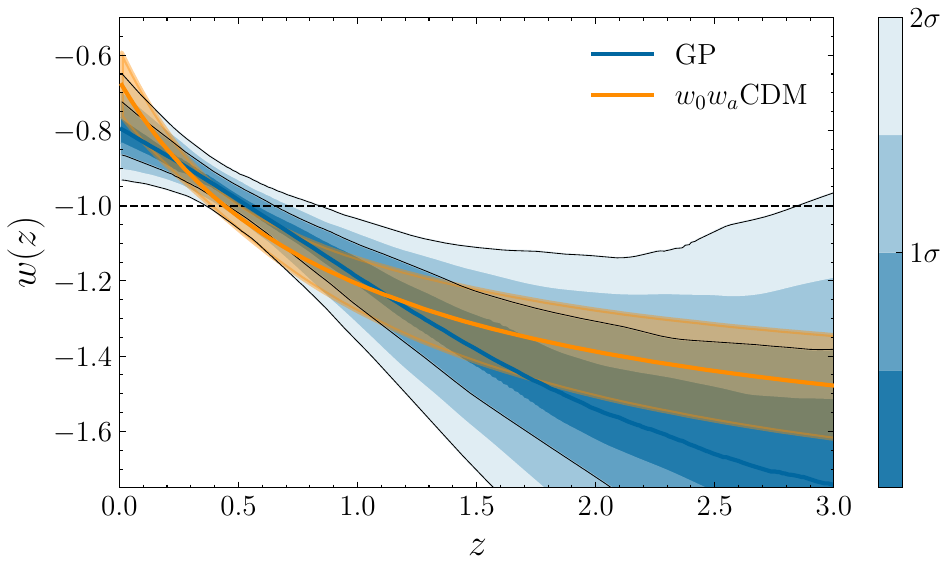}
    \end{minipage}

    \begin{minipage}{\textwidth}
        \includegraphics[width=\textwidth]{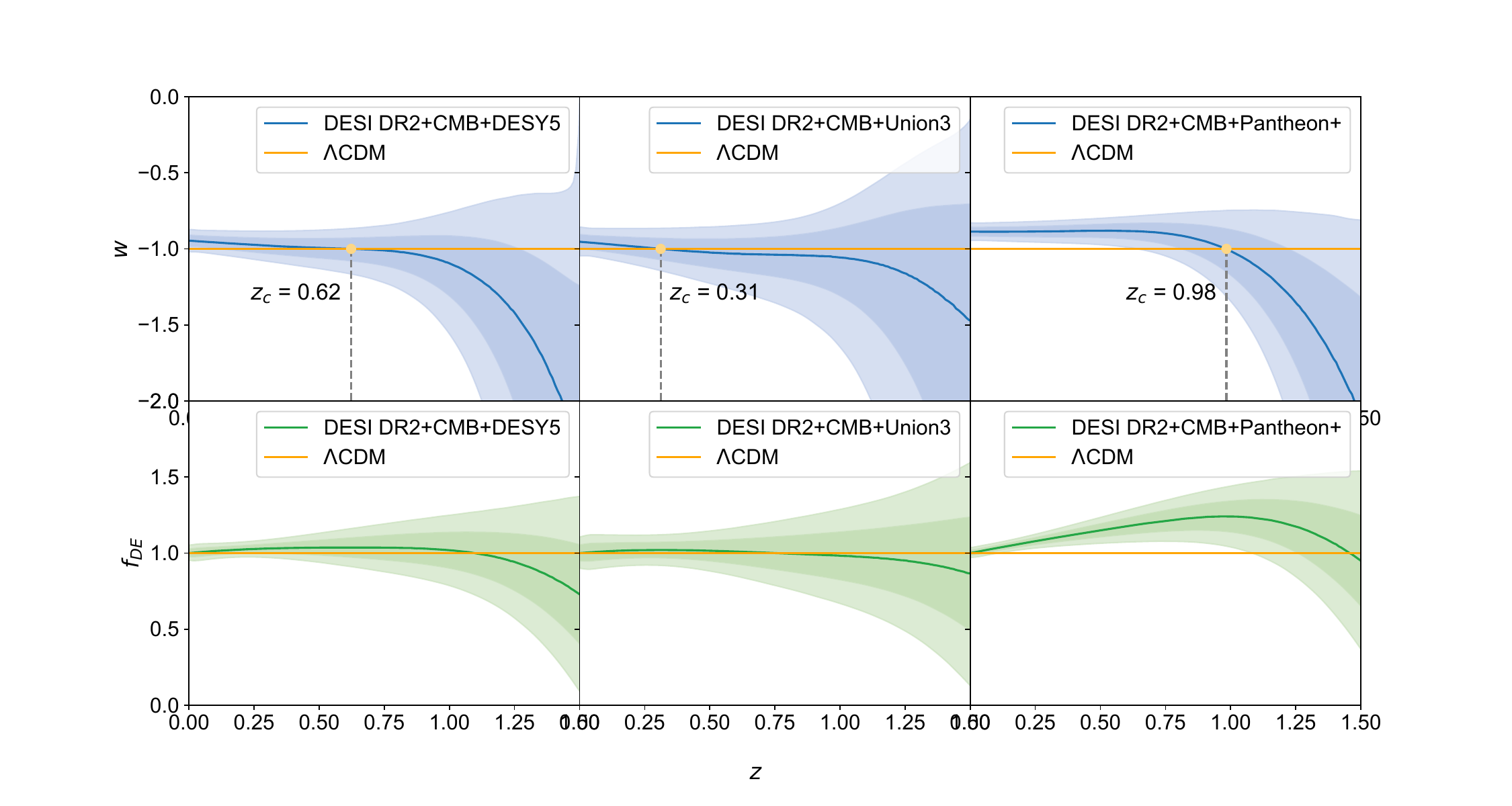}
    \end{minipage}
    \caption{\it{The reconstructed EoS parameter $w$ from Gaussian process regression. The upper left panel is from Fig.2 of~\cite{Yang:2024kdo}, the black curve denotes the mean value, while the light blue shaded zones indicate the allowed regions at 68\% confidence interval. The upper right panel is from Fig.10 of \cite{DESI:2025fii}, the Gaussian process reconstruction is shown in blue, accompanied by shaded 68\% and 95\% confidence intervals. The bottom panel is  from Fig.3 of \cite{Yang:2025mws}, shows the mean values of the reconstructed dark energy EoS parameter $w$ and the normalized energy density $f_{de}$, along with $1\sigma$ and $2\sigma$ uncertainties, where $f_{de}$ is defined as $f_{de}(z)=\rho_{de}(z)/\rho_{de,0}$.}}
    \label{fig:desi_gp}
\end{figure*}

{Last year, the BAO measurements from DESI suggested dynamical dark energy with $2.5$--$3.9\sigma$ confidence when combined with SNe datasets~\cite{DESI:2024mwx}, which generating considerable attention and discussion~\cite{Cortes:2024lgw, DESI:2024aqx, DESI:2024kob,Wang:2024dka, Giare:2024gpk, Shlivko:2024llw, Ye:2024ywg, Tada:2024znt, Carloni:2024zpl, Gialamas:2024lyw, Luongo:2024fww, Giare:2024smz, Mukherjee:2024ryz, Jiang:2024xnu, Dinda:2024kjf, Yang:2024kdo, Yang:2025kgc, Giare:2024oil, Liu:2024gfy, Wang:2024hks, Park:2024vrw, Bhattacharya:2024hep, Thompson:2024nxf, Reboucas:2024smm, Andriot:2024jsh, Orchard:2024bve, Pang:2024qyh, Colgain:2024xqj, Colgain:2024mtg, Ishak:2024jhs, Zheng:2024qzi, Escamilla-Rivera:2024sae, Yin:2024hba, Chudaykin:2024gol, Huang:2025som,RoyChoudhury:2024wri,Wolf:2024stt,Wolf:2025jlc}. The latest DESI Data Release~2, combined with supernova constraints, strengthens this preference up to $4.2\sigma$~\cite{DESI:2025zgx, DESI:2025fii,DESI:2025wyn}, motivated further investigations into dynamical dark energy~\cite{Li:2025cxn, Pang:2025lvh, Ormondroyd:2025iaf, Anchordoqui:2025fgz,   Scherer:2025esj}. Subsequent work has proceeded comprehensive observational analyses with various cosmological datasets and phenomenological parameterizations~\cite{Ling:2025lmw, Paliathanasis:2025cuc,Kessler:2025kju,Nesseris:2025lke,Ozulker:2025ehg,Silva:2025twg}. Intriguingly, the DESI data favor a Quintom behavior~\cite{Feng:2004ad}, where the dark energy EoS parameter crosses the cosmological constant boundary $w = -1$ from below. On the other hand, a wide range of theoretical models beyond the cosmological constant have been studied with the DESI observations to explain this dynamical evolution of dark energy~\cite{Chaussidon:2025npr,Luciano:2025elo}, such as various extended scalar-field models~\cite{Wolf:2025jed,Gialamas:2025pwv,Lin:2025gne,Goh:2025upc,Shajib:2025tpd}, modified gravity~\cite{Yang:2025mws,Odintsov:2025jfq,Nojiri:2025low,Paliathanasis:2025hjw,Maurya:2025hmy} and interacting dark energy scenarios~\cite{Pan:2025qwy,Silva:2025hxw,Li:2025ula,Chakraborty:2025syu}.}

We present an overview of the cosmological constraints on the $w_0w_a$ dark energy model based on DESI observations. The constraints derived from the two DESI data releases are summarized in Fig.~\ref{fig:desi_w0wa_all}. For clarity, we have added red lines to divide the parameter space into four distinct regions, corresponding to quintessence, phantom, Quintom-A, and Quintom-B. The origin marks the $\Lambda$CDM model. As shown in Fig.~\ref{fig:desi_w0wa_all}, the $\Lambda$CDM model lies well outside the $2\sigma$ credible region allowed by the posterior constraints, with both data releases favoring the Quintom-B regime.

{The CPL parametrization fixed the higher-order terms of Taylor expansion. This kind of parameterization is just an effective low-redshift parameterization, does not introduce any extra information from the higher order terms. To examine the possible higher order effects and avoid such a bias, one could extend the parametrization to include higher-order terms or adopt data-driven approaches~\cite{Nesseris:2025lke}.}

Thus, non-parametric approaches are very important and necessary. One widely used non-parametric method is the Gaussian process regression \cite{Shafieloo:2012ht, Holsclaw:2010sk, Holsclaw:2010nb, Seikel:2012uu}. The Gaussian-process regression allows one to reconstruct the function and its derivatives in a model-independent manner from observational data points, utilizing a chosen kernel covariance function. The results derived from the Gaussian process regression are shown in Fig.~\ref{fig:desi_gp}. The results from this non-parametric approach are consistent with those from the $w_0w_a$ parameterization.

\begin{figure*}[htbp]
    \centering
        \includegraphics[width=\textwidth]{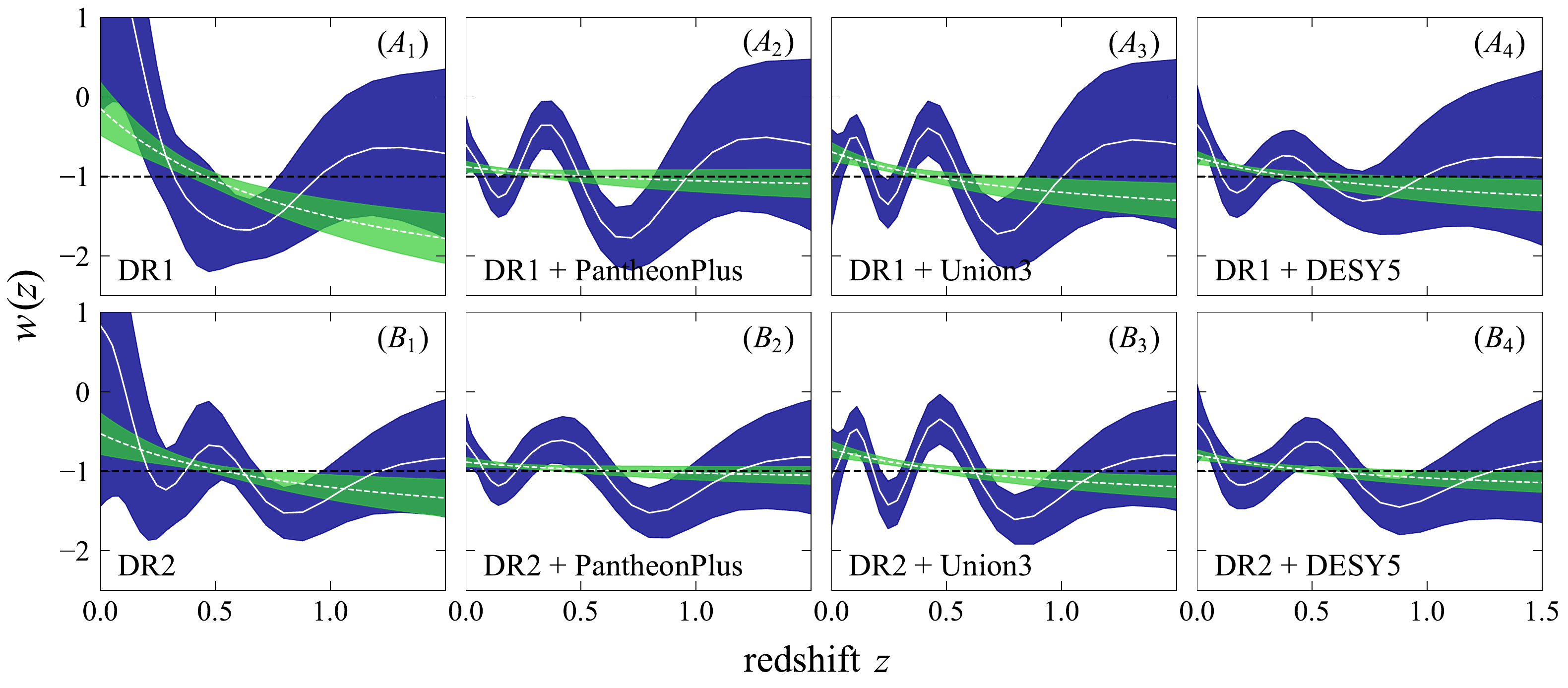}
    \caption{\it{ Dark energy equation of state $w(z)$ reconstructed from several datasets. The results are for two different approaches: the correlation-prior method (bottom-layered, dark-blue band) and the ($w_0$, $w_a$) parameterization (top-layered, green band).The panel is  from Fig.3 of ~\cite{DESI:2025wyn}. }}
    \label{fig:na_w}
\end{figure*}

In Ref.~\cite{DESI:2025wyn}, the DESI collaboration performed a non-parametric Bayesian reconstruction of $w(z)$ with Principal Component Analysis ~\cite{Huterer:2002hy} by jointly analyzing DESI BAO, SNe, and CMB data shown in Fig.~\ref{fig:na_w}, finding consistent results with companion DESI papers. {The result of reconstructed $w(z)$CDM models showed the significance of $w\neq-1$ reaches $4.3 \sigma$ for DESI DR2 BAO + DESY5, $3.9 \sigma$ for DESI DR2 BAO + Union3 and $3.1 \sigma$ for DESI DR2 BAO + PantheonPlus, favoring dynamical dark energy with EoS crossing $-1$. Using the same method, this preference for dynamical dark energy has been found in 2012
with a significance of $2.5 \sigma$ from the combined data using SNLS3 and weak prior~\cite{Zhao:2012aw} and in 2017 with a significance of $3.5 \sigma$ from the combined dataset ALL16~\cite{Zhao:2017cud}. Different parameterization and non-parametric reconstructions with different data combination including DESI data all yield a Quintom dark energy evolutionary behavior, demonstrating the robustness of this behavior~\cite{DESI:2025wyn,Li:2025vuh}.}

\section{QUINTOM MODELS AND NO-GO THEOREM}
\label{sec:4}

When examining the manifestations of Quintom dark energy, it is crucial to revisit the No-Go theorem associated with dynamical dark energy. As reviewed in~\cite{Cai:2009zp} (see also \cite{Qiu:2010ux}), the basic single-field models or single perfect fluid models cannot exhibit Quintom behavior due to the No-Go theorem, which strictly prohibits the EoS parameter $w$ from crossing the cosmological constant boundary in such simple frameworks \cite{Vikman:2004dc, Deffayet:2010qz}. Generally, we treat dark energy as an additional non-interacting fluid (along with matter and radiation). In conformal Newtonian gauge $ds^2=-a(\eta)^2 [ -(1+2\Psi) {\rm d} \eta^2+(1-2\Phi)\delta_{ij} {\rm d} x^i {\rm d} x^j ]$ with no anisotropic perturbations (leads to $\Psi=\Phi$), the dark energy perturbation in the Fourier space can be described as \cite{Kodama:1984ziu,Ma:1995ey}
\begin{align}
    \delta '&=- \Big (1+\frac{\bar{p}}{\bar{\rho}}\Big)(\theta-3\Phi')-3\mathcal{H} \Big (\frac{\delta p}{ \delta \rho}-\frac{\bar{p}}{\bar{\rho}} \Big )\delta~, \label{eq:delta perturbation} \\
    \theta'&=-\Big(\mathcal{H}+\frac{\bar{p}'}{\bar{\rho}+\bar{p}}\Big)\theta+ k^2 \Big ( \frac{\delta p}{\bar{\rho}+\bar{p}} + \Psi \Big)~,  \label{eq:theta perturbation}
\end{align}
where the prime denotes the derivative with respect to conformal time $\eta$, satisfying $ad\eta=dt$, $\mathcal{H}$ represents the conformal Hubble parameter and 
$\delta \equiv \delta \rho / \bar{\rho}$, $\theta \equiv ik^{j}\delta T^{0}_{j}/(\bar{\rho}+\bar{p})$ are the density contrast and velocity perturbations respectively.

One can consider a barotropic perfect fluid characterized by its pressure $\bar{p}$, energy density $\bar{\rho}$, and EoS parameter $w=\bar{p}/\bar{\rho}$. The adiabatic sound speed for such a fluid can be expressed as:
\begin{equation}
    c_a^2=\frac{\delta p}{\delta \rho}\Big |_{\rm{adiabatic}}=\frac{\bar{p}'}{\bar{\rho}'}=w-\frac{w'}{3\mathcal{H}(1+w)}.
\end{equation}
It should be noted that when $w$ crosses $-1$, the expression for the sound speed diverges, leading to unphysical instabilities in dark energy perturbations.

If the fluid is non-barotropic, the perturbation of dark energy will naturally lead to the variation of the entropy. In such a case, the dark energy perturbation will lead to non-adiabatic (isocurvature) perturbations. A more general definition of the sound speed should come from the relation between the gauge invariant pressure $\delta \hat{p}$ and energy density $\delta \hat{\rho}$,
\begin{align}
    \delta \hat{p}&=\delta p+3\mathcal{H}c_a^2(1+w)\bar{\rho}\frac{\theta}{k^2}, \\
    \delta \hat{\rho}&=\delta \rho+3\mathcal{H}(1+w)\bar{\rho}\frac{\theta}{k^2}.
\end{align}
In a rest frame, the gauge invariant terms $\delta \hat{p}$ and $\delta \hat{\rho}$ are coincident with the pressure $p$ and energy density $\rho$ of the fluid, thus we define
\begin{equation}
    c_s^2=\frac{\delta \hat{p}}{\delta \hat{\rho}}=\frac{\delta p}{\delta \rho} \Big |_{\rm{rest-frame}}.
\end{equation}
Consequently, we derive the relation between $\delta p$ and $\delta \rho$ in a general frame as
\begin{equation}
    \delta p = c_s^2 \delta \rho +(c_s^2-c_a^2) \Big[ 3 \mathcal{H}(1+w)\bar{\rho} \Big]\frac{\theta}{k^2}.
\end{equation}
Substituting this relation into Eq.~\eqref{eq:theta perturbation}, we get
\begin{equation}
\begin{aligned}
\theta'&=-\Big [\mathcal{H}-3\mathcal{H}(c_s^2-c_a^2+w)+\frac{w'}{1+w} \Big ]\theta\\ &+k^2 \Big (\frac{c_s^2}{1+w}\delta + \Psi \Big) \\
 &=-\mathcal{H}(1-3w)\theta+k^2 \Psi\\ &+\frac{1}{1+w}\Big[3\mathcal{H}(1+w)(c_s^2-c_a^2)\theta-w'\theta+k^2 c_s^2\delta \Big ] \\
 &=-\mathcal{H}\theta+k^2\Psi+\frac{k^2\delta \hat{p}}{(1+w)\bar{\rho}}.
\end{aligned}
\label{eq:non-ad v-perturbation}
\end{equation}

From the definition of the velocity perturbations and Eq.~\eqref{eq:non-ad v-perturbation}, we can see that $\theta$ and $\theta'$ will be divergent when crossing the cosmological constant boundary, unless we have $\delta \hat{p}=0$ when crossing. The gauge invariant entropy perturbation $\hat{\Gamma}$ is described as
\begin{equation}
    \hat{\Gamma}=\frac{1}{w\bar{\rho}}(\delta p -c_a^2 \delta \rho)=\frac{1}{w \bar{\rho}}(\delta \hat{p}-c_a^2\delta \hat{\rho}).
\end{equation}
Due to the divergence of the adiabatic sound speed $c_a^2$ at the crossing point, to obtain a finite value of $\hat{\Gamma}$, we have to set $\delta \hat{\rho}=0$. However, in such a condition, we naturally have $\delta p = c_a^2\delta \rho$, which contradicts the non-adiabatic perturbations we assumed. Thus, a smooth crossing in non-adiabatic perturbations is not allowed. In conclusion, we can summarize that for a single perfect fluid, it is impossible to realize $w$ crossing $-1$.

Analogously, this divergence phenomenon extends to generic single scalar field without higher derivatives. In general, the No-Go theorem states: in the Friedmann-Robertson-Walker (FRW) universe described by a single perfect fluid or a single scalar field $\phi$ with a Lagrangian $L = L(\phi, \partial_\mu \phi\partial^\mu\phi)$, which minimally couples to Einstein gravity, its equation of state $w$ cannot cross the cosmological constant boundary~\cite{Xia:2007km,Zhao:2005vj,Cai:2009zp,Qiu:2010ux}.

\section{MODEL BUILDING OF QUINTOM DARK ENERGY}
\label{sec:5}

This No-Go theorem explicitly showed that, to realize the Quintom scenario, one ought to include more degrees of freedom either via higher derivative of the matter field, or more fields {to derive a health dispersion relation when crossing the cosmological boundary}, or modified gravity {which regards dark energy as an effective description of scenario beyond general relativity}. In the following, we will demonstrate several examples of Quintom dark energy.

\subsection{Multi-field model}

The first model of Quintom dark energy was proposed in April of 2004 \cite{Feng:2004ad} which combined a quintessence field $\phi$ with a phantom field $\sigma$ as shown below, 
\begin{equation}\small
 S=\int d^4 x \sqrt{-g}\left[-\frac{1}{2} \nabla_\mu \phi \nabla^\mu \phi+\frac{1}{2} \nabla_\mu \sigma \nabla^\mu \sigma-V(\phi, \sigma)\right]
\end{equation}
Thus, the effective energy density $\rho$ and the effective pressure $p$ are given by
\begin{equation}\small
\rho=\frac{1}{2} \dot{\phi}^2-\frac{1}{2} \dot{\sigma}^2+V(\phi, \sigma),\quad  p=\frac{1}{2} \dot{\phi}^2-\frac{1}{2} \dot{\sigma}^2-V(\phi, \sigma),
\end{equation}
and the corresponding EoS is now given by
\begin{equation}
w=\frac{p}{\rho}=\frac{\dot{\phi}^2-\dot{\sigma}^2-2 V(\phi, \sigma)}{\dot{\phi}^2-\dot{\sigma}^2+2 V(\phi, \sigma)}
\end{equation}

There are also other variations of double field quintom model~\cite{Hu:2004kh,Wei:2005nw,Chimento:2008ws,Saridakis:2009jq}. Besides, a generalized multiple scalar field model $\phi_i$ can be given by~\cite{Mughal:2020glg,Vazquez:2023kyx}
\begin{equation}\footnotesize
S=\int d^4 x \sqrt{-g}\left[-\frac{1}{2} \sum_i \epsilon_i  \nabla_\mu \phi_i \nabla^\mu \phi_i-V({\phi_1,\phi_2,...\phi_n}) \right]
\end{equation}
where $\epsilon_i=\left\{\begin{aligned}
    &+1,\quad \text{Quintessence}\\
    &-1, \quad \text{Phantom}
\end{aligned} \right.$, for the distinction between quintessence and phantom fields.

The associating quantities are given by
 \begin{align}
&\rho_{\rm total}=\frac{1}{2} \sum_i \epsilon_i \dot{\phi}_i^2+V({\phi_1,\phi_2,....,\phi_n}), \\&  p_{\rm total}=\frac{1}{2} \sum_i \epsilon_i \dot{\phi}_i^2-V({\phi_1,\phi_2,....,\phi_n}), \\& w_{\rm total}=\frac{\sum_i \epsilon_i \dot{\phi}_i^2-2 V({\phi_1,\phi_2,....,\phi_n})}{\sum_i \epsilon_i \dot{\phi}_i^2+2  V({\phi_1,\phi_2,....,\phi_n})}
\end{align}


\subsection{Single scalar field with higher derivatives}

A single scalar Quintom model with higher derivatives can be found in \cite{Li:2005fm, Zhang:2006ck, Cai:2007gs}. In DHOST \cite{Langlois:2017mxy, Langlois:2018jdg} and Horndeski \cite{Horndeski:1974wa} as well as Galileon \cite{Nicolis:2008in, Deffayet:2009wt, Deffayet:2011gz} theories, the Lagrangian is delicately designed to have a coupling between the kinetic term, $X \equiv \partial_\mu\phi\partial^\mu\phi/2$ and the higher derivative $\Box\phi$, so that the model can have its EoS across $-1$ without pathologies such as ghost instabilities  \cite{Matsumoto:2017qil, Tiwari:2024gzo}.

Specifically, we take the model with Lagrangian as \cite{Li:2011qfa}:
\begin{equation}
    L= -X+c_1X\Box\phi+c_2X\phi^2~,
    \label{lagrangian}
\end{equation}
where $c_{1,2}$ are constants. From the Lagrangian one gets its pressure and energy density as: 
\begin{eqnarray}
    \rho&=&(c_2\phi^2-1+6c_1H\dot\phi)X~,\\
    p&=&(c_2\phi^2-1-2c_1\ddot\phi)X~,
\end{eqnarray}
Thus the EoS is
\begin{equation}
    w\equiv\frac{p}{\rho}=\frac{c_2\phi^2-1-2c_1\ddot\phi}{c_2\phi^2-1+6c_1H\dot\phi}~,
    \label{eos}
\end{equation}
where dot denotes derivative on cosmic time $t$. However, in order to compare with the data, it is convenient to transfer the variable from $t$ to the scale factor, $a$. Thus, Eq. \eqref{eos} becomes:
\begin{equation}\small
\begin{aligned}
    \label{eos2}
    &w\equiv\frac{p}{\rho}\\ &=\frac{c_2\phi^2-1-c_1(2aH^2\phi'+a^2(H^2)'\phi+2a^2H^2\phi'')}{c_2\phi^2-1+6c_1aH^2\phi'}~,      
\end{aligned}
\end{equation}
where prime denotes derivative on $a$. 

The dynamics of this model is analyzed in \cite{Li:2011qfa}. One can set the initial condition of $\phi$ to be $\phi_i=0$ at $a_i$, and then in order to ensure the positivity of the energy density, there is $6c_1aH^2\phi'>1$. Moreover, following the analysis in \cite{Li:2011qfa}, the velocity of $\phi$ can be set as $\phi'=A/aH^2$, therefore in matter-dominant era where $(H^2)'=-3H^2/a$, one has \begin{equation}
    w\simeq \frac{-1-3c_1A}{-1+6c_1A}~.
\end{equation}
In \cite{Li:2011qfa} where $c_1A\rightarrow 1/3$, $w\rightarrow -2$. On the other hand, a positive $\phi'$ makes $\phi$ increase, and when the $c_2\phi^2$ term in both $p$ and $\rho$ dominates over other terms, the equation of state $w$ will reach $1$. As a result, it is very natural to have $w$ cross $-1$ in this model.

In order to compare with the recent DESI data, we need to expand the equation of state around the point $a=1$. From Eq. \eqref{eos2} with $\phi'=A/aH^2$, one has:

\begin{equation}\small
    \begin{aligned}
    &\phi(a)=\int \frac{Ada}{aH^2}\\ &\simeq\frac{Ae^{6w_{T0}}}{2}(a-a_i)[2(a+a_i-1)+3w_{T0}(a+a_i-2)]~,\label{eqw1} 
    \end{aligned}
\end{equation}

{where $w_{T0}$ is the total equation of state of the universe today. Then $w$ could be expressed as the function of $A,c_1,c_2,a_i$. Compared with the CPL parametrization of $w$, i.e. $w=w_0+w_a(1-a)$, the corresponding function of $w_0$ and $w_a$ can be derived.
Consequently, we can inversely solve for the coefficients $c_{1,2}$, expressing them in terms of $w_0$ and $w_a$.}

\begin{figure*}[ht]
    \centering
    \includegraphics[width=0.8\linewidth]{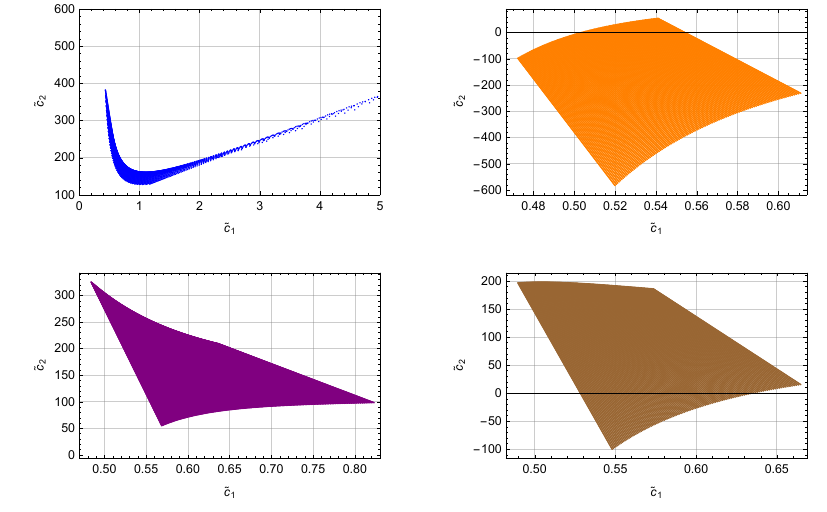}
    \caption{The constraint on coefficients $\tilde{c}_1$ and $\tilde{c}_2$ from DESI+CMB data (blue), DESI+CMB+Pantheon+ data (orange), DESI+CMB+Union3 data (purple), DESI+CMB+DESY5 data (brown). We choose $a_i=0.5$ corresponding the redshift $z_i\simeq 1$ which is in the matter-dominant era. The total equation of state $w_{T0}\simeq w_0\Omega_{DE0}$, where $\Omega_0$ is the fraction of today's dark energy density, $\Omega_{DE0}\simeq0.647$.}
    \label{constrainG}
\end{figure*}

The DESI data gives constraints on $w_0$ and $w_a$: $w_0=-0.42\pm0.21$, $w_a=-1.75\pm0.58$ (DESI+CMB), $w_0=-0.838\pm0.055$, $w_a=-0.62^{+0.22}_{-0.19}$ (DESI+CMB+Pantheon+), $w_0=-0.667\pm0.088$, $w_a=-1.09^{+0.31}_{-0.27}$ (DESI+CMB+Union3), $w_0=-0.752\pm0.057$, $w_a=-0.86^{+0.23}_{-0.20}$ (DESI+CMB+DESY5) \cite{DESI:2025zgx}. Using the observations, we can further constrain the model parameters $c_{1,2}$. Numerical constraints are performed in Fig. \ref{constrainG}. Note that $c_{1,2}$ actually has degeneracy with the overall factor of $\phi$ field, namely $A$, so we are practically constraining the combined coefficients: $\tilde{c}_1\equiv c_1A$, $\tilde {c}_2\equiv c_2A^2$. 

    \begin{figure*}[htbp]
  \centering
  \begin{minipage}[b]{0.48\linewidth}
    \includegraphics[width=1\linewidth]{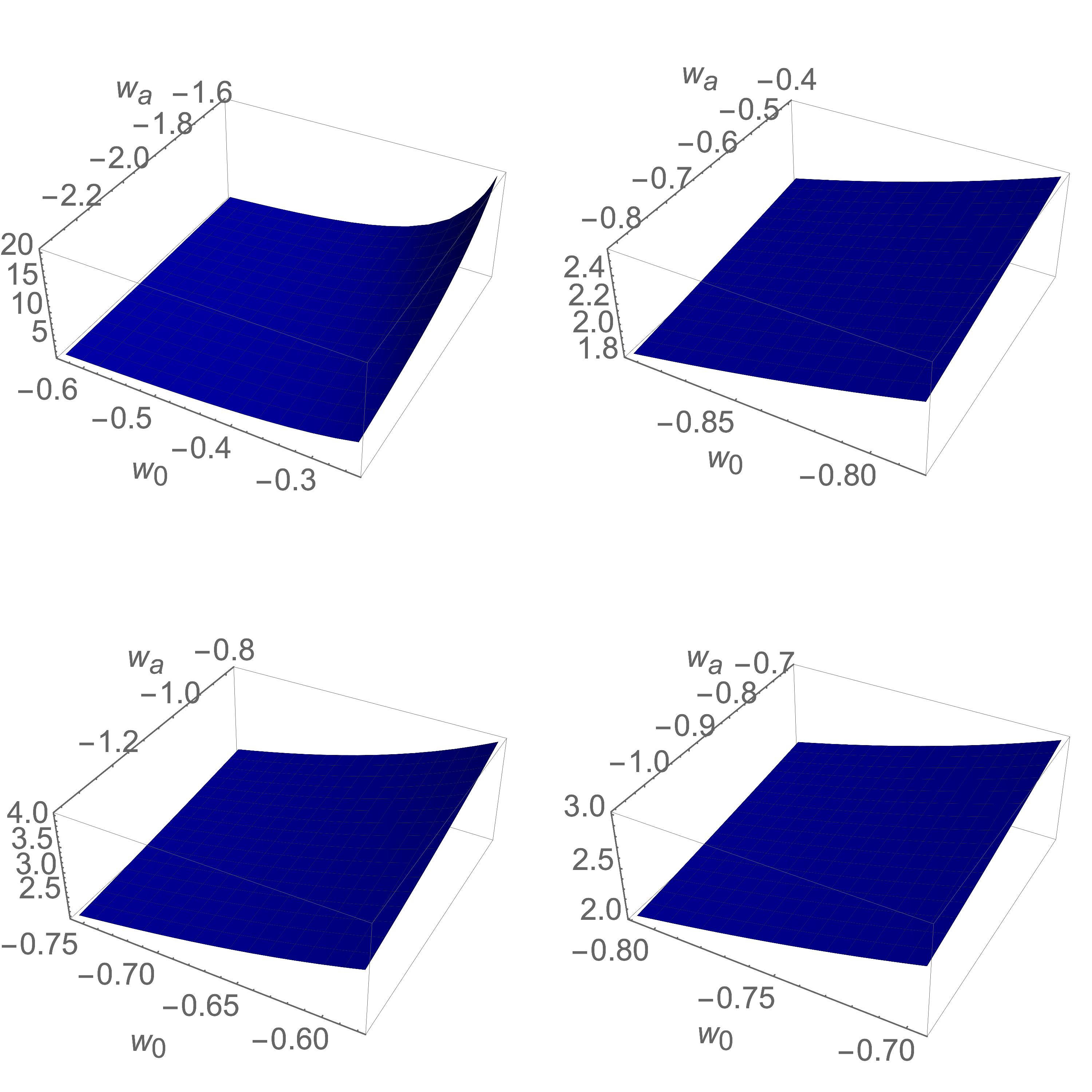}
    \caption{The value of factor $D$ (vertical axis) in terms of $w_0$ and $w_a$ in the allowed parameter space of DESI+CMB (left-top), DESI+CMB+Pantheon+ data (right-top), DESI+CMB+Union3 data (left-bottom), DESI+CMB+DESY5 data (right-bottom). The parameter choices are the same with Fig. \ref{constrainG}.}
        \label{constrainD}
          \end{minipage}
  \hfill
  \begin{minipage}[b]{0.48\linewidth}
    \centering
    \includegraphics[width=1\linewidth]{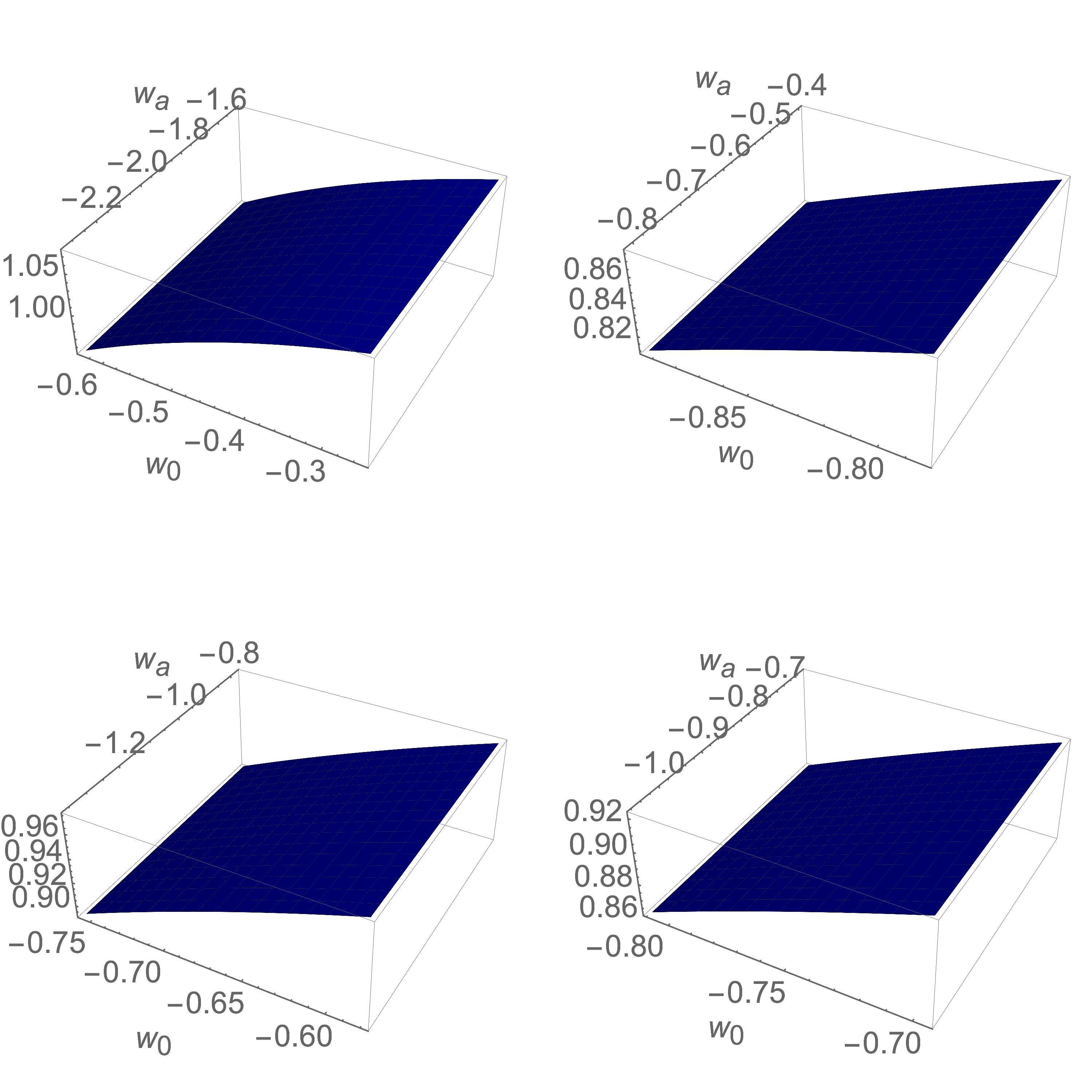}
    \caption{The value of factor $c_s^2$ (vertical axis) in terms of $w_0$ and $w_a$ in the allowed parameter space of DESI+CMB (left-top), DESI+CMB+Pantheon+ data (right-top), DESI+CMB+Union3 data (left-bottom), DESI+CMB+DESY5 data (right-bottom). The parameter choices are the same with Fig. \ref{constrainG}.}
    \label{constraincs}
    \label{fig:constraincs_mp}
  \end{minipage}
\end{figure*}

The Figure \ref{constrainG} shows the parametric region of $\{\tilde{c}_1,\tilde{c}_2\}$ which is constrained by datasets of DESI+CMB, DESI+CMB+Pantheon+, DESI+CMB+Union3 and DESI+CMB+DESY5 data. One can see that there is large parameter space in this model to fit the current DESI DR2 data. Moreover, one can also check the perturbations of this model to see that there are no pathologies of instabilities. The perturbed Lagrangian from Eq. \eqref{lagrangian} turns out to be \cite{Li:2011qfa}:
\begin{equation}
    \delta L\sim D [(\partial_t{\delta\phi})^2-a^{-2}c_s^2(\partial_i\delta\phi)^2]~,
\end{equation}
where $\delta\phi$ is the perturbation of $\phi$ field, and
\begin{eqnarray}
    D&=&c_2\phi^2-1+6c_1H\dot\phi~, \\
    c_s^2&=&\frac{c_2\phi^2-1+2c_1(\ddot\phi+2H\dot\phi)}{c_2\phi^2-1+6c_1H\dot\phi}~.
\end{eqnarray}
It is obvious that if $D>0$, there will be no ghost instability, and if $c_s^2>0$, there will be no gradient instability. 

Although it is difficult to analyse their positivity in an analytical approach, using the ansatz solution $\phi'=A/aH^2$ as well as function of $c_{1,2}$, we can express $D$ and $c_s^2$ in terms of $w_0$ and $w_a$, and we plot them with respect to  $w_0$ and $w_a$ within the allowed region by DESI DR2 data, which are presented in Figs. \ref{constrainD} and \ref{constraincs}. From the Figures we can see that, for all the allowed regions, we always have $D>0$ and $c_s^2>0$, which shows that our model is healthy without any instabilities.


\subsection{Modified gravity models}

Besides matter field approaches, the modified gravity (MG) approach is another way to produce dark energy effect and thus the Quintom scenario \cite{Starobinsky:1980te, Hehl:1994ue, Gronwald:1997bx, JimenezCano:2021rlu, Aoki:2023sum, Capozziello:2002rd, Cai:2015emx, Krssak:2018ywd, Joyce:2016vqv, Clifton:2011jh, BeltranJimenez:2017tkd, Heisenberg:2023lru,Wu:2024vcr}, bringing dark energy a gravitational interpretation. A common-considered framework is metric-affine gravity (MAG)~\cite{Hehl:1994ue,Gronwald:1997bx,JimenezCano:2021rlu}, which describes gravity with a metric and a general affine connection. In this formalism, a general affine connection $\Gamma^{\alpha}_{\ \mu\nu}$ can be decomposed
as
\begin{equation}
 \Gamma^{\alpha}_{\ \mu\nu} = \mathring\Gamma^{\alpha}_{\ \mu\nu}  
+L^{\alpha}_{\ \mu\nu}+K^{\alpha}_{\ \mu\nu},
\end{equation}
where $\mathring\Gamma^{\alpha}_{\ \mu\nu}$ is the Levi-Civita connection, $L ^{\alpha}_{\ \mu\nu}$ and $K^{\alpha}_{\ \mu\nu}$ are the disformation tensor and contortion tensor respectively, characterizing the deviation of 
the full affine connection from the Levi-Civita one. The general affine connection yields the curvature tensors, torsion tensor and non-metricity tensor as~\cite{Heisenberg:2023lru}

\begin{align}\small
R^{\sigma}{}_{\rho\mu\nu} &\equiv \partial_{\mu} \Gamma^{\sigma}{}_{\nu\rho}\! - 
 \!
 \partial_{\nu} \Gamma^{\sigma}{}_{\mu\rho}\!\\& +\! \Gamma^{\alpha}{}_{\nu\rho} 
 \Gamma^{\sigma}{}_{\mu\alpha} \!- \!\Gamma^{\alpha}{}_{\mu\rho} 
 \Gamma^{\sigma}{}_{\nu\alpha} ~, \nonumber \\ \nonumber
 T^{\lambda}{}_{\mu\nu} &\equiv
 \!\Gamma^{\lambda}{}_{\nu\mu}-\Gamma^{\lambda}{}_{\mu\nu}\! ~,\\  \nonumber
 Q_{\rho \mu \nu} &\equiv \nabla_{\rho} g_{\mu\nu} = \partial_\rho g_{\mu\nu} - \Gamma^\beta{}_{\rho\mu} g_{\beta\nu} - \Gamma^\beta{}_{\rho\nu} g_{\mu\beta}  ~.
\end{align}
 {Curvature causes parallel transport along a closed curve to change the vector being transported. Torsion is the anti-symmetric part of the connection, which refers to the asymmetry of parallel transport when exchanging the transported vector and direction of transport. Non-metricity describes how the metric changes under parallel transport, leading to changes in the length of the vector.} The Ricci scalar $R$ in MAG can be written in terms of the Ricci scalar corresponding to the Levi-Civita connection as \cite{Bahamonde:2021gfp,CANTATA:2021asi}
\begin{equation}
    R =\mathring R-Q+T+C+B.
    \label{eq:ricci scalar}
\end{equation}
Where the non-metricity scalar $Q$, torsion scalar $T$, mixing scalar $C$ and boundary term $B$ are given by
\begin{equation}
    \begin{aligned}
     Q=&\frac{1}{4}Q^{\alpha}Q_{\alpha}- 
\frac{1}{2}\tilde{Q}^{\alpha}Q_{\alpha}\\&-\frac{1}{4}Q_{\alpha\mu\nu}Q^{
\alpha\mu\nu}+\frac{1}{2}Q_{\alpha\mu\nu}Q^{\nu\mu\alpha}, \\
    T=&-T^{\tau}T_{\tau}+\frac{1}{4}T_{\rho\mu\tau}T^{\rho\mu\tau}+\frac{1}{2}T_{\rho\mu\tau}T^{\tau\mu\rho}, \\
    C=&\tilde{Q}_{\rho}T^{\rho}-Q_{\rho}T^{\rho} +Q_{\rho\mu\nu}T^{\nu\rho\mu}, \\
    B=&\mathring\nabla_{\rho}\Big (Q^{\rho}-\tilde{Q}^{\rho}+2T^{\rho}\Big ),
    \end{aligned}
\label{eq:mag scalar}
\end{equation}
with $Q_{\alpha}=g^{\mu\nu}Q_{\alpha\mu\nu}$ and $\tilde 
Q_{\alpha}=g^{\mu\nu}Q_{\mu\alpha\nu}$  representing the two independent traces of 
the non-metricity tensor, and $T^{\mu}=T^{\nu\mu}{}_{\nu}$ is the trace of torsion tensor. The general action of MAG can be constructed as an arbitrary function of the fundamental geometric quantities of curvature, torsion, and non-metricity.
Such a general formulation can reduce to $f(R)$~\cite{Starobinsky:1980te, Capozziello:2002rd, Nojiri:2010wj}, $f(T)$~\cite{Cai:2015emx, Bahamonde:2021gfp, Krssak:2018ywd, Krssak:2015oua, Aldrovandi:2013wha}, and $f(Q)$ gravity~\cite{BeltranJimenez:2017tkd, BeltranJimenez:2019tme, Heisenberg:2023lru, Koussour:2023ulc} under certain conditions, based only on curvature, torsion or non-metricity respectively. The action for these modified gravity theories can be uniformly expressed as  
\begin{align}
S=\int d^4 x \sqrt{-g}\left[\frac{1}{16\pi G} f(X)+\mathcal{L}_{\mathrm{m}}\right]
\end{align}
where $X$ represents $R, T$ or $Q$, with $R, T, Q$ the Ricci scalar, torsion scalar and non-metricity scalar, $\mathcal{L}_{\mathrm{m}}$ represents the matter Lagrangian density respectively. Under the flat  FRW metric $\mathrm{d} s^2=\mathrm{d} t^2-a(t)^2\left(\mathrm{~d} r^2+r^2 \mathrm{~d} \theta^2+r^2 \sin ^2 \theta \mathrm{~d} \phi^2\right)$, with $a(t)$ the scale factor. The modified Friedmann equations can be expressed effectively as

\begin{align}
3 H^2 & =\rho_{\mathrm{m}}+\rho_{\mathrm{de}} \\
-2 \dot{H}-3 H^2 & =p_{\mathrm{m}}+p_{\mathrm{de}}
\end{align}

where $\rho_{\mathrm{m}}$ and $p_{\mathrm{m}}$ denote the energy density and pressure of matter, and the effective energy density $\rho_{\text {de }}$ and pressure $p_{\text {de }}$ are in terms of the gravitational modifications.

In $f(R)$ gravity, we have

\begin{align}
\rho_{\mathrm{de}, R}= & \frac{1}{f_R}\left[\frac{1}{2}\left(f-R f_R\right)-3 H \dot{R} f_{R R}\right] \\
p_{\mathrm{de}, R}= & \frac{1}{f_R}\left(2 H \dot{R} f_{R R}+\ddot{R} f_{R R}\right) \\  +&\frac{1}{f_R}\left[\dot{R}^2 f_{R R R}-\frac{1}{2}\left(f-R f_R\right)\right]
\end{align}

where $R=-12 H^2-6 \dot{H}$ and $f_R=\mathrm{d} f / \mathrm{d} R, f_{R R}=$ $\mathrm{d}^2 f / \mathrm{d} R^2$, and accordingly the effective dark-energy EoS is $w \equiv p_{\mathrm{de}, \mathrm{R}} / \rho_{\mathrm{de}, \mathrm{R}}$.

Among these three mostly considered modified gravity models, $f(R)$ models often yield higher-order field equations and potential instabilities, requiring careful construction. At the same time, $f(T), f(Q)$ only contains second order equations.

In $f(T)$ gravity with the form of $f(T)=T+F(T)$, the effective energy density and pressure of additional gravitational part as dark energy could be written as

\begin{align}
\rho_{\mathrm{de}, \mathrm{T}} & =-\frac{1}{2} F+T F_T \\
p_{\mathrm{de}, \mathrm{T}} & =\frac{F-T F_T+2 T^2 F_{T T}}{2+2 F_T+4 T F_{T T}}
\end{align}

with $T=-6 H^2$. thus the effective EoS parameter of dark energy is $w \equiv p_{\mathrm{de}, \mathrm{T}} / \rho_{\mathrm{de}, \mathrm{T}}$. 

The $f(Q)$ gravity within the coincident gauge in the FRW metric has the same evolution at the background level, where $Q=-6 H^2$. The corresponding expressions of the coincident gauge $f(Q)$ case can be obtained from the one of $f(T)$ gravity, with replacement $T \rightarrow Q$. 

{Meanwhile, in order to acquire a healthy $f(T)$ theory, we also need no instabilities and this requires
\begin{equation}
    \Omega^2=\frac{-\frac{f}{4}-T^2f_{TT}}{f_T+2T f_{TT}} \geq 0.
\end{equation}
For instance, a model with a negative $\Omega^2$ is obviously unstable against gravitational perturbations.}

For specific functional forms of $f$, these modified gravity models can yield Quintom dark energy behavior consistent with observational data~\cite{Ren:2021tfi,Briffa:2020qli,Yang:2024tkw,Yang:2025kgc,Basilakos:2025olm}.

\subsection{The Effective Field Theory Approach}

The effective field theory (EFT) approach provides a powerful and unified way to study different dynamical dark energy models~\cite{Gubitosi:2012hu, Bloomfield:2012ff}.According to this concept, the extra scalar degree of freedom that results from spontaneous symmetry breaking of the time translation in an expanding universe emerges as a Goldstone boson~\cite{Cheung:2007st, Piazza:2013coa}. It is notable that the EFT framework can incorporate both single-scalar field theories and modified gravity theories, such as $f(R)$ gravity and Horndeski theory within curvature-based EFT~\cite{Bloomfield:2012ff, Tsujikawa:2014mba}, as well as $f(T)$ gravity in torsion-based EFT~\cite{Li:2018ixg, Yan:2019gbw, Ren:2022aeo}. This unified framework makes it possible to conduct systematic comparisons and studies of different modified gravitational theories under a common theoretical structure.

The most general EFT action in metric-affine gravity can be written as \cite{Yang:2025mws}
\begin{equation}
    \begin{aligned}
    S=&\int d^4x\sqrt{-g}\Bigg[ \frac{M_P^2}{2}\Big (\Psi(t)\mathring R+d(t)T+e(t)Q\Big ) \\+&\frac{M_P^2}{2}\Big (g(t)T^0+h(t)Q^0+j(t)\tilde{Q}^0 \Big)\\
    -&\Lambda(t)-b(t)g^{00}-k(t)Q^{000} +\frac{M_P^2}{2}m(t)C \Bigg]\\+&S_{DE}^{(2)},
    \end{aligned}
    \label{eq:EFT MAG}
\end{equation}
with $M_p^2=1/8\pi G$  the Planck mass, and where $\Psi$, $\Lambda$, $d$, $e$, $g$, $h$, $j$, $b$, $k$ and $m$ are functions of the time coordinate $t$. Additionally, 
$S_{DE}^{(2)}$  contains all operators from the perturbation level.

In the curvature-based case, the EFT form can be simplified to
\begin{equation}
\begin{aligned}
    S=&\int \mathrm{d}^4x \sqrt{-g}\left[ \frac{M_p^2}{2}\Psi (t)R-\Lambda(t)-c(t)g^{00}\right]\\+&S_{DE}^{(2)},
    \label{eq:eft action}
     \end{aligned}
\end{equation}

$\Psi$ indicates whether the scalar field is minimally coupled. Therefore, we can define the effective density and pressure of dark energy in EFT frame as
\begin{align}
    \rho_{de}^{eff}&=\frac{1-\Psi}{\Psi}\rho_m-3M_p^2H\frac{\dot{\Psi}}{\Psi}+\frac{c}{\Psi}+\frac{\Lambda}{\Psi}  \\
    p_{de}^{eff}&=\frac{1-\Psi}{\Psi}p_m+M_p^2\frac{\Ddot{\Psi}}{\Psi}+2M_p^2H\frac{\dot{\Psi}}{\Psi}+\frac{c}{\Psi}-\frac{\Lambda}{\Psi}.
\end{align}
Then, the effective dark energy EoS parameter of the general EFT form can be expressed as
\begin{equation}\small
    \begin{aligned}
    &w_{de}^{eff}=\frac{(1-\Psi)p_m+M_p^2\Ddot{\Psi}+2M_p^2H\dot{\Psi}+c-\Lambda}{(1-\Psi)\rho_m-3M_p^2H\dot{\Psi}+c+\Lambda} \\
    &=-1+\frac{(1-\Psi)(\rho_m+p_m)+M_p^2\Ddot{\Psi}-M_p^2H\dot{\Psi}+2c}{(1-\Psi)\rho_m-3M_p^2H\dot{\Psi}+c+\Lambda}.
    \end{aligned}
\end{equation}

The action \eqref{eq:eft action} could represent the minimally coupled single-field dark energy model when $\Psi (t)=1$. The standard quintessence scenario has the Lagrangian in the unitary gauge as
\begin{equation}
    -\frac{1}{2}(\partial \phi)^2-V(\phi) \stackrel{unitary}{\longrightarrow}-\frac{1}{2}\dot{\phi}_0^2(t)g^{00}-V(\phi_0),
\end{equation}
when corresponding to the background action of \eqref{eq:eft action}, we obtain
\begin{equation}
    \Psi (t)=1, \quad c(t)=\frac{1}{2}\dot{\phi}_0^2(t), \quad \Lambda(t)=V(\phi_0).
\end{equation}

In the case of $f(R)$ gravity the action can be rewritten in the unitary gauge by choosing the background value $R^{(0)}=t$ as
\begin{equation}
    f(R) \stackrel{unitary}{\longrightarrow} f_R(R^{(0)})R+f(R^{(0)})-R^{(0)}f_R(R^{(0)}),
\end{equation}
then we have
\begin{equation}
\begin{aligned}
    \Psi (t)&=f_R(R^{(0)}), \quad c(t)=0, \\
    \Lambda(t)&=-\frac{M_p^2}{2}\Big (f(R^{(0)})-R^{(0)}f_R(R^{(0)}) \Big ).
\end{aligned}
\end{equation}

In $f(T)$ gravity \cite{Cai:2015emx}  the geometry is flat and metric-compatible, and thus in the unitary gauge we have
\begin{equation}
    f(T) \stackrel{unitary}{\longrightarrow}  f_T(T^{(0)})T+f(T^{(0)})-f_T(T^{(0)})T^{(0)}.
\end{equation}
the non-zero terms can be obtained by comparing with Eq.~\eqref{eq:EFT MAG}
\begin{equation}
    \begin{aligned}
        \Psi(t)&=-f_T(T^{(0)}), \quad d(t)=2\dot{f_T}(T^{(0)}) \\
        \Lambda(t)&=-\frac{M_p^2}{2}\Big[ f(T^{(0)})-T^{(0)}f_T(T^{(0)}) \Big ],
    \end{aligned}
\end{equation}

Similar to the steps of $f(T)$ case, the $f(Q)$ action in the unitary gauge has the form as
\begin{equation}
    f(Q)\stackrel{unitary}{\longrightarrow} f_Q(Q^{(0)})Q+f(Q^{(0)})-f_Q(Q^{(0)})Q^{(0)}.
\end{equation}
The corresponding non-zero terms are
\begin{equation}
    \begin{aligned}
        &\Psi(t)=f_Q(Q^{(0)}), \quad j(t)=-h(t)=\Dot{f_Q}(Q^{(0)}), \\
        &\Lambda(t)=-\frac{M_p^2}{2}\Big [f(Q^{(0)})-Q^{(0)}f_Q(Q^{(0)}) \Big ].
    \end{aligned}
\end{equation}
In particular, the background evolution for $f(Q)$ cosmology in the coincident gauge is identical to that for $f(T)$ gravity.

\subsection{Interacting dark energy}

{Along the line of the Quintom scenario with two fields, one has considered a system involving a scalar field with neutrinos~\cite{Zhang:2005eg, Zhao:2009ke} or dark matter \cite{Wang:2005jx, Wang:2005ph, Wang:2016lxa, Yang:2018uae, Yang:2018euj, Andriot:2025los,Linder:2025zxb} and also interacting two-fluid model~\cite{Zhang:2005kj}. }

{The interacting dark energy model is also an important candidate for explaining DESI observations. It describes the existence of interactions between dark energy and dark matter, leading to energy transfer between them. This results in an additional interaction term in their conservation equations, generally denoted as $Q_{\mathrm{int}}$.} 

\begin{align}
\dot{\rho}_{\mathrm{de}}+3 H(1+w) \rho_{\mathrm{de}}=Q_{\mathrm{int}}, \\
\dot{\rho}_{\mathrm{dm}}+3 H \rho_{\mathrm{dm}}=-Q_{\mathrm{int}},
\end{align}

{The interaction term $Q_{\mathrm{int}}$ is  often parameterized as a function of the Hubble constant, dark matter energy density, and dark energy density, such as $\xi H\rho_{\mathrm{dm}}$, $\xi H\rho_{\mathrm{de}}$ and other possible forms~\cite{DiValentino:2019ffd,Pan:2019gop}. The scenario of interacting dark energy can also be derived from a Lagrangian perspective. By considering the coupling between a scalar field dark energy $\phi$ and a spinor field dark matter $\psi$, the corresponding interaction term $Q_{\mathrm{int}}(\phi,\psi)$ can also be obtained~\cite{Khoury:2025txd, Wang:2025znm}. }

{Due to the presence of the interaction, this part can be regarded as contributing to the effective equation of state of dark energy, effectively altering the properties of the dark sectors. Of course, its impact on cosmic evolution also depends on the form of the interaction. Appropriate interaction forms can make the effective dark energy equation of state consistent with the quintom-like behavior from DESI result~\cite{Li:2024qso,Pan:2025qwy,Silva:2025hxw,Li:2025ula,Chakraborty:2025syu}.}

\subsection{Compare different models}

{The Quintom behavior is a result based on observations. This indicates that the effective dynamical dark energy component extending beyond the $\Lambda$CDM framework must exhibit an EoS parameter that crosses the $-1$. Moreover, the Quintom characteristics also imply that dark energy is unlikely to be a simple cosmological constant or a canonical single scalar field but rather has a more complex evolutionary dynamics.}

{In order to break through the conditions discussed in No-Go theorem and achieve the evolution of Quintom, various models have introduced different approaches. Based on single scalar field dark energy models, multi-scalar models incorporate additional fields as degrees of freedom. But crossing the $w=-1$ boundary necessarily requires a phantom field with a negative kinetic term and violates NEC. This kind of ghost field suffers from quantum instabilities\cite{Carroll:2003st, Cline:2003gs}. }

{Single scalar field models with higher derivatives introduce coupling between the kinetic term and higher derivative terms. The presence of this coupling term allows the scalar field to exhibit richer evolutionary behavior, thereby enabling the realization of a quintom model. However, general higher derivative models still retain ghost modes. To obtain a stable theory, requirements must be imposed on the coupling term, such as in degenerate higher-derivative models~\cite{Matsumoto:2017qil, Tiwari:2024gzo,Li:2011qfa}.}

{Modified gravity theories approach the problem from a different perspective by extending the gravitational action of general relativity. The additional gravitational terms can be regarded as an effective dynamic dark energy component, driving the accelerated expansion of the universe. Since the effective equation of state parameter is an effective representation of the gravitational terms not directly related to the physical energy density, $w_{eff}$ can naturally cross the boundary of $-1$ without physically violating the NEC. Modified gravity theories typically introduce new degrees of freedom, which also alter the evolution of perturbations. The specific forms of modified gravity models also need to satisfy the corresponding stability conditions, and the behavior of the extra degrees of freedom remains an active area of research~\cite{Bahamonde:2021gfp,Heisenberg:2023lru}.}

{Interacting dark energy models introduce dynamical behavior for dark energy by incorporating a coupling between dark energy and dark matter. This not only affects the dynamics of dark energy but also alters the evolution of dark matter. The effective quintom behavior is also closely dependent on the form of the interaction term. Meanwhile, it is necessary to constrain the form or parameter space of the interaction term to ensure its impact on large-scale structure aligns with observations and maintains a stable interactive system~\cite{DiValentino:2019ffd,Li:2024qso}.}

There are also approaches studying the realization of the Quintom scenario via fermion fields \cite{Cai:2008gk, Wang:2009ae, Wang:2009ag, Dil:2016vod}, holographic dark energy model\cite{Zhang:2005hs,Zhang:2006qu,Nojiri:2005pu,Luciano:2025elo, Wu:2025vfs}. Some studies have approached DESI result
from the perspective of dark matter, such as a non-zero equation of state parameter for dark matter\cite{Kumar:2025etf, Wang:2025zri, Chen:2025wwn, Li:2025eqh}, as well as many other mechanisms that can be found in the literature. 
 
{According to above statement, a Quintom-like evolution of $w(z)$ can be reproduced by many different microphysical theoretical frameworks, which is highly degenerate at the background level. It is difficult to distinguish these models only on the basis of observations at the level of cosmological background evolution. To break this degeneracy and  identify the most likely theoretical explanation for quintom dark energy, it is necessary to incorporate the behavior of perturbations. Employing effective field theory, we can investigate the perturbation-level properties characteristic of various model categories.}  

{Modified gravity models not only influence the evolution of the cosmic background but also affect structure formation, gravitational waves, gravitational potentials, and other related phenomena. Meanwhile, interacting dark energy models impact physical processes associated with dark matter. Models involving interactions with photons will also produce effects on CMB polarization, which will be discussed in detail in the following chapter. A future multi-messenger observations at  perturbation level combined growth of structure, gravitational lensing and slip, gravitational waves, and CMB polarization provide tests and constraints on dynamical dark energy scenarios, offering deeper insights into the fundamental nature of quintom dark energy.}

\section{IMPLICATIONS OF QUINTOM COSMOLOGY}
\label{sec:6}

\subsection{Interactions of Quintom Dark Energy with ordinary matter}

Being a dynamical field, Quintom dark energy is expected to couple directly to the ordinary matter in the universe. In general, one imposes the shift symmetry, $\phi\rightarrow \phi+c$, for these interactions. The shift symmetry demands that the dark energy scalar can only have derivative couplings to matter fields. 
At the leading order, the derivative couplings have the form $\partial_{\mu}\phi J^{\mu}$ and produce spin-dependent forces. These forces between microscopic particles cannot superpose into long range forces between unpolarized macroscopic objects. In addition, the shift symmetry prevents large radiation corrections.

The couplings to the baryon current $\partial_{\mu}\phi J_B^{\mu}$ or to the $B-L$ current $\partial_{\mu}\phi J_{B-L}^{\mu}$ can be used to construct the baryogenesis or letpogenesis models~\cite{Li:2001st,Li:2002wd}, in which the baryon number asymmetry observed today was produced at thermal equilibrium in the early universe. During the evolution of the Quintom dark energy, $\dot \phi$ does not vanish, these coupling terms violate the Lorentz and CPT symmetries. This explains why in these models the baryon number asymmetry was produced thermally. 

The dynamic dark energy field couples to photons derivatively will take the form of Chern-Simons coupling,
\begin{equation}
\mathcal{L}_{CS}=\frac{c}{M}\partial_{\mu}\phi A_{\nu}\tilde{F}^{\mu\nu}~,
\end{equation}
where $c$ is the coupling constant, $M$ is mass scale from the viewpoint of effective field theory, $\tilde{F}^{\mu\nu}=(1/2)\epsilon^{\mu\nu\rho\sigma}F_{\rho\sigma}$ is the dual of the electromagnetic field tensor.
Through the Chern-Simons coupling, the evolution of the quintom field in the universe induces CPT violation in the photon sector and can be potentially observed by CMB polarization experiments. 
For single light, the polarization direction of the photon got rotated under the Chern-Simons coupling when the photon transported from the source to the observer. This rotation expressed in terms of the Stokes parameters as 
\begin{equation}
(Q\pm iU)'=\exp (\pm i2\chi)(Q\pm iU)
\end{equation}
and the rotation angle depends on the field difference  \cite{Li:2008tma}, 
\begin{equation}
\chi =\frac{c}{M}\Delta \phi\equiv \frac{c}{M}[\phi(x_s)-\phi(x_o)]~.
\end{equation}
For CMB photons, the source is the last scattering surface, $x_s=x_{lss}$. This in turn changes the CMB power spectra, especially produces non-vanishing $TB$ and $EB$ correlations \cite{Lue:1998mq}, here $T, E, B$ refer to the temperature, the E polarization, the B mode polarization respectively. By assuming an isotropic rotation angle $\chi=\bar\chi$, one may obtain the full set of equations for the rotated CMB spectra \cite{Feng:2006dp}
\begin{eqnarray}
& &{C'}_l^{TE}=C_l^{TE}\cos 2\bar\chi~,\nonumber\\
& &{C'}_l^{TB}=C_l^{TE}\sin 2\bar\chi~,\nonumber\\
& &{C'}_l^{EE}=C_l^{EE}\cos^2 2\bar\chi+C_l^{BB}\sin^2 2\bar\chi~,\nonumber\\
& &{C'}_l^{BB}=C_l^{EE}\sin^2 2\bar\chi+C_l^{BB}\cos^2 2\bar\chi~,\nonumber\\
& &{C'}_l^{EB}=\frac{1}{2}(C_l^{EE}-C_l^{BB})\sin 4\bar\chi~.
\end{eqnarray}
The first detection of the CMB rotation angle was performed in Ref. \cite{Feng:2006dp}, where a nonzero rotation angle $\bar \chi =-6.0\pm 4.0~{\rm deg}$ is mildly favored by CMB polarization data from the three-year WMAP observations and the January 2003 Antarctic flight of BOOMERanG. 

Again, due to the dynamical nature of Quintom dark energy, the rotation angle cannot be isotropic absolutely. 
The Quintom field fluctuates spatially as well as evolves in the time direction. Its inhomogeneous distribution on the last scattering surface will induce the anisotropies of the rotation angle  \cite{Li:2008tma}, $\chi=\bar\chi +\delta\chi$. According to $\chi =\frac{c}{M}[\phi(x_s)-\phi(x_o)]$, the anisotropic rotation angle $\delta \chi=\frac{c}{M}\delta\phi(x_{lss})$ depends on the perturbation of the Quintom field around the last scattering surface. This brings further distortions to the CMB spectra and is potentially observable to the future CMB experiments  \cite{Li:2008tma, Li:2013vga, Li:2017drr,Zhai:2025hqt}.

\subsection{Quintom Cosmology in very early Universe }

In standard cosmological frameworks such as the $\Lambda$CDM model and conventional inflationary scenarios, the evolution of the early universe inevitably traces back to an initial singularity -- an extreme state of divergent curvature, energy density, and temperature where general relativity ceases to be valid. To resolve this fundamental issue, alternative models have been proposed, among which bounce cosmology offers a promising route to non-singular early-universe evolution \cite{Starobinsky:1980te, PhysRevLett.68.1969, Peter:2002cn, Piao:2003zm, Biswas:2005qr, Creminelli:2007aq, Cai:2007qw, Wei:2007rp, Novello:2008ra, Cai:2014bea}. In this section, we demonstrate that non-singular models can emerge naturally within the framework of Quintom cosmology.

Bounce cosmology encompasses scenarios in which the universe initially undergoes a phase of contraction, reaches a finite minimum scale (the bounce), and subsequently enters an expanding phase. Near the bounce, the momentary violation of NEC becomes essential. To smoothly transition from the contracting to the expanding phase and eventually match the hot Big Bang conditions, the EoS parameter must evolve from $w<-1$ to $w>-1$.

This behavior can be understood by examining the dynamics of the scale factor $a(t)$ and the Hubble parameter $H(t)$. During contraction, $\dot{a} < 0$, and during expansion, $\dot{a} > 0$. At the bounce point, $\dot{a} = 0$, and a successful bounce requires $\ddot{a} > 0$ in its neighborhood. Equivalently, the Hubble parameter transitions from $H < 0$ to $H > 0$, passing through $H = 0$ at the bounce, which implies $w<-1$ in its neighborhood. However, to avoid a subsequent Big Rip -- as seen in purely phantom dark energy models -- the universe must evolve toward a standard thermal history, necessitating a transition from $w < -1$ to $w > -1$. This Quintom-like evolution is thus not only compatible with bounce cosmology but also a requirement for its viability. We can also obtain the bouncing solution in Quintom models with a phenomenological realization \cite{Cai:2007qw}. Beyond phenomenological constructions, bounce cosmology can also be implemented using fundamental field-theoretic models. Models of this class have been studied in \cite{Piao:2003zm, Cai:2007qw, Cai:2008qb, Cai:2008qw}, and the dynamics of their perturbations have been analyzed in \cite{Cai:2007zv, Cai:2008ed, Cai:2009fn, Cai:2009rd, Cai:2009hc, Karouby:2011wj, Alexander:2014uaa, Addazi:2016rnz}.

Moreover, within the same theoretical framework, it is possible to construct cyclic cosmologies, where a bounce experiences a turn-around phase during expansion, it triggers contraction and subsequently generates the next bounce, resulting in an oscillatory cyclic. The idea of a cyclic universe \cite{1934rtc..book.....T}, has since been revisited in various contexts, including higher dimensional string theory \cite{Khoury:2001wf, Steinhardt:2002ih, Khoury:2001bz, Buchbinder:2007ad} and loop quantum cosmology \cite{Zhang:2007bi, Xiong:2007cn, Wei:2007rp, Banijamali:2012ph}. An appealing aspect of cyclic Quintom cosmology is its ability to avoid both the Big Rip and Big Crunch singularities while naturally incorporating periodic phases of acceleration. Furthermore, since the scale factor grows from cycle to cycle, the model predicts a progressively flatter universe, addressing the flatness problem without fine-tuning. One such cyclic model based on a parameterized Quintom EoS has also been proposed to alleviate the coincidence problem \cite{Feng:2004ff}.

The emergent-universe scenario posits a past-eternal cosmos with a finite nonzero scale factor as $t\to -\infty$, thus avoiding the Big Bang singularity \cite{Ellis:2002we}. In its original form, the universe is asymptotically Einstein static before entering an inflationary phases \cite{Ellis:2003qz}. Quintom fields allow such an emergent phase even in flat FRW model \cite{Cai:2013rna, Ilyas:2020zcb}. In these cases, the analytic solutions demonstrate a smooth transition from a quasi-static initial phase to radiation-like expansion, with no curvature singularity encountered.

Additionally, there are also many other non-singular models of the very early universe motivated by nonconventional theories. For instance, adding higher-derivative “ghost condensate” terms to the action allows a smooth bounce in ekpyrotic models \cite{Lehners:2015mxa}. It was found that fermion condensation of the Nambu-Jona-Lasinio type can also induce a non-singular bouncing solution \cite{Tukhashvili:2023itb}. In beyond-Horndeski scalar-tensor theories, one can design a non-singular bounce as well \cite{Kolevatov:2017kvx}.

\section{SUMMARY AND FUTURE PROSPECT}\label{conclusion}

{In this paper, we have briefly reviewed the Quintom models of dark energy. 
We review the historical development of dark energy, from the static cosmological constant paradigm toward the observationally supported dynamical behavior. 
When dark energy EoS $w(z)$ deviates from $-1$, its phenomenology can be broadly categorized into quintessence, phantom and Quintom. 
The latest DESI DR2 results combined with complementary cosmological datasets provide substantive evidence for dynamic dark energy.  CPL parametrization and non-parametric approaches consistently indicate that the dark energy EoS parameter $w(z)$ evolves across $-1$, which corresponds to a Quintom-B scenario.}

{We have shown that due to the No-Go theorem, the Quintom model needs more degrees of freedom. Multi-field scalar theories, higher-derivative single-field frameworks, modified gravity and interacting dark energy could realize the Quintom scenario and serve as candidate theories for dark energy under the current observational trend. An effective field theory framework could bridge the scalar field and geometric approaches, providing a powerful and unified way to study different dynamical dark energy models.} 

{We have considered the interactions of the Quintom field with ordinary matter, and pointed out that these interactions will result in spin-dependent forces and generate the matter and anti-matter asymmetry in the universe. The Quintom interaction with the photon will cause the rotation of the CMB polarization. This can be served as a further test of Quintom models of dark energy, in addition to the measurements on EoS.}

{Forthcoming DESI data releases with enhanced statistical accumulation and extended redshift coverage are expected to yield a deep understanding of the essential nature and dynamical evolution of dark energy. Euclid and Rubin could provide a cross-check of DESI result with  complementary measurements of BAO. Future supernova surveys such as the ZTF and the Rubin Observatory will supplement critical constraints on the low redshift universe. Meanwhile, next-generation CMB experiments including the Simons Observatory, CMB-S4, and AliCPT will refine our understanding of the early universe and tighten constraints on dynamical dark energy models. Notably, precise measurements of the CMB TB and EB power spectra in the future CMB polarization observation could serve as a probe for a possible Chern-Simons coupling between dark energy and photons, offering a novel test for dynamical dark energy scenarios.}

\section{ACKNOWLEDGEMENTS}
This work was supported in part by the National Key R$\&$D Program of China Grant (2021YFC2203100, 2020YFC2201600, 2020YFC2201601 and 2021YFC2203102).

\bibliography{Quintom}{}
\bibliographystyle{elsarticle-num}

\end{document}